\documentclass[article]{aastex}
\usepackage{psfig}

\begin{document}
  \title{About the measurements of the hard X-ray background}
  \author{G.S. Bisnovatyi-Kogan  and A.S. Pozanenko  }
  \affil{  Space Research Institute (IKI), Moscow, Russia, Profsoyuznaya
  84/32, Moscow 117810, Russia}
  \email{gkogan@iki.rssi.ru}
  \begin{abstract}

  We analyze uncertainties in the cosmic X-ray background measurements performed
  by the INTEGRAL observatory. We find that the most important
  effect limiting the accuracy of the measurements is related to the intrinsic background
  variation in detectors. Taking into account
  all of the uncertainties arising  during the measurements we
  conclude that the X-ray background intensity obtained in the INTEGRAL
  observations is compatible with the historic X-ray background
  observations performed  by the HEAO-1 satellite.

  \end{abstract}

\smallskip



\section{Introduction}

Since the discovery of the cosmic X-ray background (CXB) in the
early X-ray observations (Giacconi et al., 1962) the importance of
the CXB measurement was recognized. The CXB can be considered as
an indicator of the activity of unresolved active galactic nuclei
(AGN). The spectral shape of the CXB and its normalization near
CXB peak (20-40~keV) is particularly important for estimation of
the population of Compton thick AGN, which are difficult to
resolve at lower energies, because of their high intrinsic
absorption (Ueda et al., 2003; Gilli et al., 2007; Sazonov et al.,
2008; Treister et al., 2009).

Measurement of CXB  is a sophisticated procedure requiring a
knowledge of different factors contaminating CXB. The
omnidirectional detectors  used in the pioneering  X- and
gamma-ray observations by KONUS experiment permitted to
investigate CXB in the energy range above 30~keV (Mazets, et al.,
1974). More detailed CXB observations can be performed by aperture
telescopes. However, contaminating factors such as an intrinsic
detector background, still require either precise modelling, or
several observations  with different relative contributions of the
intrinsic background and CXB. The latter technique was used in
early measurements of the hard X-ray background in the range 13-
180 keV onboard HEAO-1 (Gruber et al., 1999).

Intrinsic  background contribution in the HEAO-1 CXB experiment
was measured with the detector aperture blocked. Subtracting the
intrinsic instrument background from the total flux, measured with
the open aperture permitted the estimate of the spectrum of the
hard X-ray background (Fig. 1-2). Sky-looking data of 224 ks total
duration, and 205 ks of observation was collected with the
aperture closed when only the intrinsic detector background was
recorded.

\begin{figure}
\centerline{\psfig{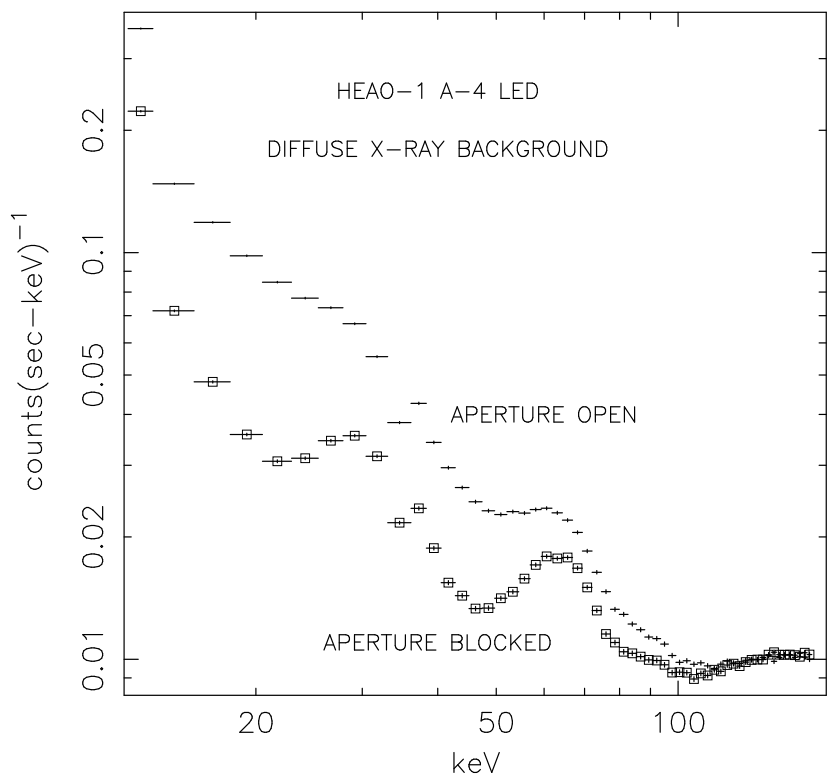}}
\caption{\label{total_net} The differential counting rates obtained
by the Low Energy Detectors on the UCSD/MIT Hard X-ray and Gamma-Ray
Instrument on the High Energy Astronomical Observatory (HEAO-1). The
difference between the rates with the detector blocked by an active
shutter, and unblocked when looking at the sky well above the
horizon, is due to the  diffuse component of the cosmic X-rays.  The
rates are averaged over the similar ranges of B,L  magnetic
coordinates. The artifact at about 32 keV is due to the energy-loss
anomaly in NaI near the K-edge.  The diffuse flux is well above the
detector background to at least 100 keV; from Gruber et al.(1999).}
\end{figure}

\begin{figure}
\centerline{\psfig{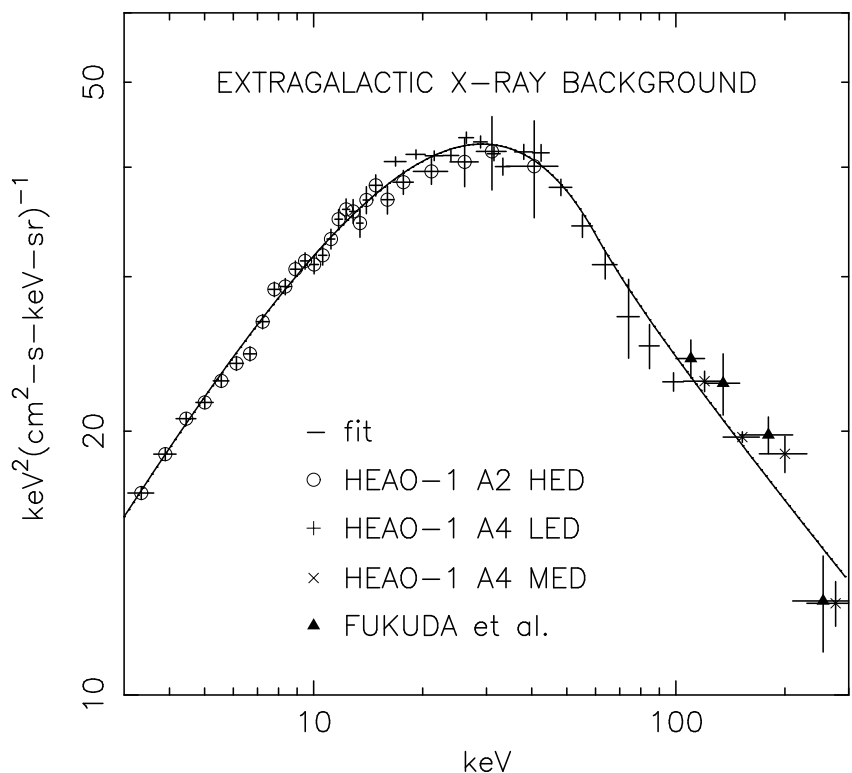}}
\caption{\label{hump} Corrected photon spectrum of the diffuse
components measured by the HEAO--1 by several different detectors,
in comparison with other data in this energy range, expressed as
spectral intensity per logarithmic energy unit dI/d($\ln$ E).  The
HEAO 1 A-4 Low Energy Detectors join smoothly to other data at
both higher and lower energies, and are in agreement with balloon
data, for which that of Fukada et al. (1975) has been chosen as
representative.  The fit is shown as a curved line.  The data from
HEAO-1 A2 were taken from High-Energy Detector no. 1. These A2
points may reflect minor artifacts of the spectral inversion; from
Gruber et al. (1999).}
\end{figure}

In the new CXB measurements performed by the INTEGRAL observatory
(Churazov et al., 2007; T\"{u}rler et al., 2010), Beppo-Sax
(Frontera et al., 2007), and BAT/Swift (Ajello et al., 2008) the
Earth was used as a natural block, and a special procedure was
used to get rid of the intrinsic instrument background, based on
the observations of CXB modulated by the Earth in FOV of the
telescopes.

In 2006 the INTEGRAL observatory (Winkler et al., 2003) performed
  four runs of $\sim$30~ksec observations with the Earth disk crossing
the field of view of the JEM-X, IBIS/ISGRI and SPI telecopes
(Churazov et al., 2007, hereafter CE2007). Indeed the Earth is not
a perfect block, and additional estimations of a non-background
X-ray emission were completed:  estimation of the Earth albedo
(i.e. the true CXB reflection by Earth atmosphere) (Churazov et
al, 2008), and modelling of the proper emission of the Earth
atmosphere resulting from a bombardment by the cosmic rays
(Sazonov et al., 2007). As a reference point the spectrum obtained
early by the HEAO-1 observations (Gruber at al, 1999) was used. No
attempt has been made to revise the spectral shape, and only the
normalization of the spectral curve was verified. The CE2007
claims that "The observed flux near the peak of the CXB spectrum
is $\sim$10\% higher than suggested by Gruber et al., 1999.", see
Fig. 3.

A different approach was used in the CXB estimations by T\"{u}rler
et al. 2010 (hereafter TM2010). The authors did not fix the
 form of the CXB spectrum, while fitting different
components (true CXB, Earth albedo, and Earth emission), using
count rate of IBIS/ISGRI during the Earth occultation
observations. Based on the same data used in CE2007 they concluded
that the CXB normalization is fully compatible with the HEAO-1
observations (Gruber at al, 1999).

\begin{figure}[int]
\centerline { \psfig{figure=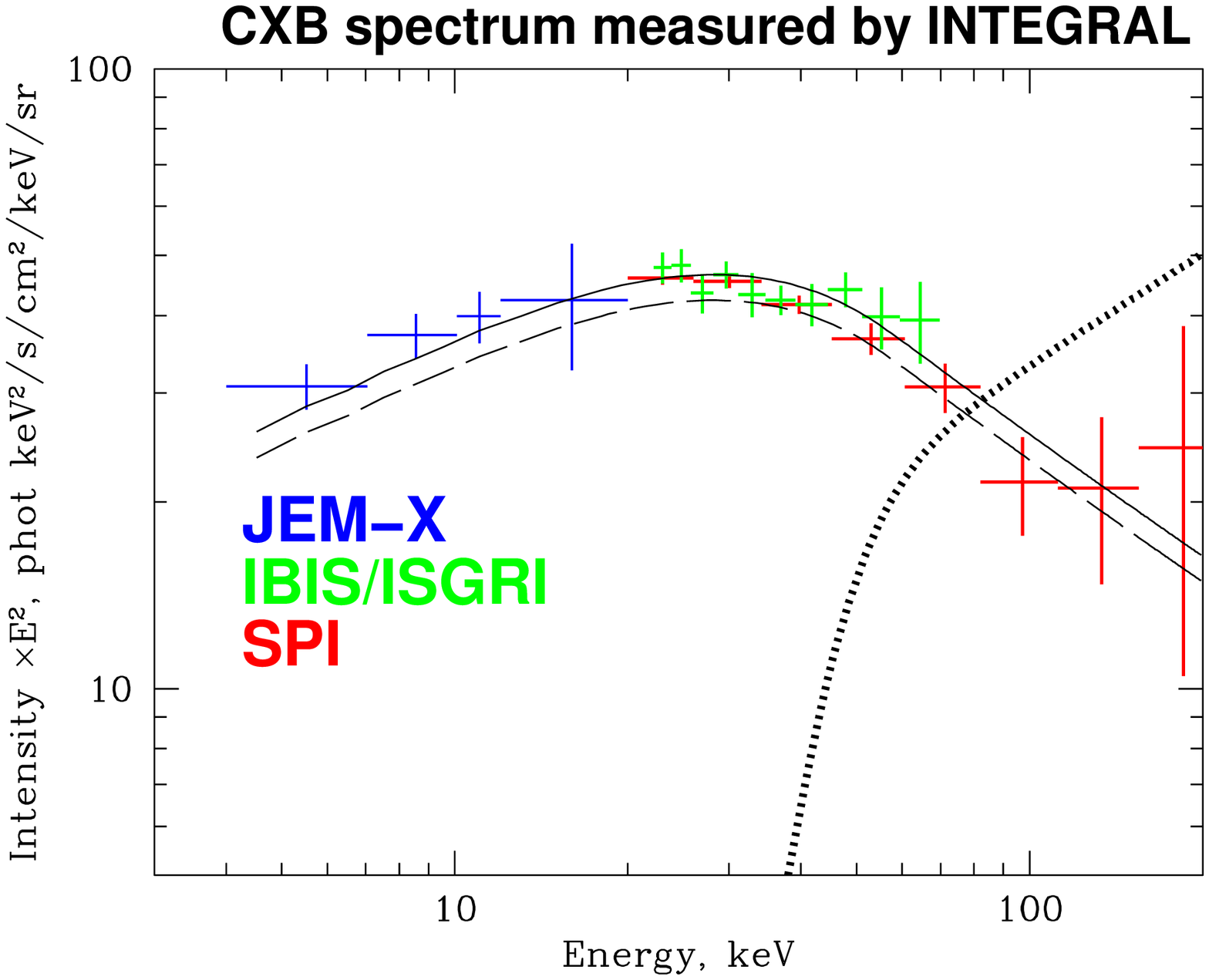,width=10cm,angle=-0} }
\caption{Spectrum of the CXB measured by INTEGRAL instruments. The
error bars plotted account for the uncertainties in the
normalization of the atmospheric emission component. The dashed
line shows the analytic approximation of the CXB spectrum by
Gruber et al. (1999). The solid line shows the same spectrum with
the best-fit normalization obtained in the work of CE2007. The
thick dotted line shows the best-fit spectrum of the Earth
atmospheric emission, from CE2007.} \label{figINT}
\end{figure}

Considering  different CXB normalization obtained in the same
INTEGRAL observations, we discuss factors which influence on the
precision of the measurement of CXB by occultation technique:  the
variability of the intrinsic background induced by the energetic
particles along the orbit of the INTEGRAL observatory,  compact
X-ray sources reflected by the Earth atmosphere, variability
induced by the non-resolved compact X-sources in the field of view
of the telescope. The last two factors have been already estimated
in CE2007, and we revise the estimates. In CE2007  the intrinsic
background in the detectors was assumed to be constant, while the
actual background level varies along the orbit and introduces an
additional uncertainty in the CXB measurement. In TM2010  the
intrinsic background variation was taken into account by simple
tracing of the IBIS/ISGRI background via SPI-ACS count rate. In
TM2010 the
 uncertainties induced by the proper Earth emission due
to the cosmic rays, as well as  specific uncertainties due to
Galactic ridge X-ray in FOV have  also  been considered.

\section{Reflection of X-ray sources emission by the Earth atmosphere}

The Earth atmosphere reflects X-rays from both CXB and compact
sources. Reflected emission contaminates measurement of CXB when
the Earth is used as a block. Indeed the CXB reflection dominates
over the Galactic sources reflection. The contribution of the
 emission from the Crab nebula (the most
intense Galactic source at 30 keV), reflected by the Earth, in
comparison with CXB, was estimated as low as $\sim$ 1\% (CE2007).
Using the same approach we analyze the contribution of other
compact X-ray sources.

Sources behind the Earth  do not give any input, and for the
source in front of the Earth the effective illuminated square is
equal to $\pi\, R_\oplus^2$. When the angle $\theta$ between the
directions to the Earth and to the source is in the range
$0<\theta<\pi$, the effective square for the reflected radiation
which  illuminates the Earth and falls onto the instrument, is
equal to $\pi(1-\cos\theta)R_\oplus^2/2$ ($\theta=0$ corresponds
to the source behind the Earth).  We analyze reflection of
different X-ray sources, in comparison with the Crab, taken into
account    the geometrical factor $(1-\cos\theta)/2$.

Using the sources in 20 - 40~keV energy band included in the BATSE
occultation catalog (Harmon et al., 2004) we estimate the relative
contribution of the reflected emission from  the sources of the
catalog in comparison with the Crab nebula. The relative
contribution of the reflected emission of Cyg X-1 to the Crab
nebula is equal to $\sim$ 545/900, and the relative (to the Crab)
contribution of all sources of the catalog with the flux larger
than 1~mCrab is equal to $\sim$ 2123/900=2.4. Therefore the total
uncertainties due to reflection of the compact sources in 20 -
40~keV energy band is $\sim$ 3.4.

In the energy band 2-10~keV similar estimates  can be done using
BeppoSax catalog  (Verrecchia et al. 2007). In this energy band
the Sco X-1 gives the main contribution into the reflected flux
from the Earth atmosphere. With account of the geometrical factor
the
 contribution of the Sco X-1 flux relative to the Crab is 3765/1864, and
for all sources in the sky this ratio is equal to 9212/1864=4.9.

Assuming the contamination by the Crab at 40~keV as 1\%, one can
estimate the contamination by the other sources as 2.4\%. Hence
total uncertainty in CXB measurement induced by compact sources at
40~keV is $\sim$3.4\%.  The contamination at the energies
$<10$~keV is significantly less. Taking the albedo of the Earth
atmosphere at 10~keV (40~keV) equals to 0.01 (0.3) (Churazov et
al., 2008), one can estimate the uncertainty induced by the
compact sources as $4.9 \times \frac{0.01}{0.3} $\% = 0.2\%. Using
the above values one should bear in  mind, that these estimates
were obtained for a model in which the spectrum of each compact
source is supposed to have the same spectral shape as the CXB
spectrum.

\section{Variability of the intrinsic background in detector}

The time dependent spectral distribution of the X-ray flux
measured by the INTEGRAL instruments, with the Earth in the field
of view, was approximated  by the following formula (CE2007),
\begin{eqnarray}
F(E,t) \approx C(E)-S_{\rm
CXB}(E)\left[1-A(E)\right]\Omega(t)+\nonumber \\S_{\rm
ATM}(E)\Omega(t) = C(E)-S_{\rm Earth}(E)\Omega(t).
  \label{eq1}
\end{eqnarray}
Here $S_{\rm ATM}(E)$ is the spectrum of the cosmic ray induced
atmospheric emission (averaged over the Earth disk and normalized
per solid unit angle), and $S_{\rm Earth}(E)=S_{\rm
CXB}(E)\left[1-A(E)\right]-S_{\rm ATM}(E)$ is the combined flux of
all components related to the presence of the Earth in the FOV,
and A(E) is the energy dependent albedo of the Earth atmosphere.
The term $C(E)$ includes the intrinsic background of the detector,
 the total combined flux of Galactic sources, and CXB. The term  $C(E)$ is considered as a
constant in the paper CE2007. We show here that the variability of
the intrinsic background induced by the energetic particles along
the orbit should introduce a most significant uncertainty in the
CXB measurement.

\subsection{Background evolution in detectors}

The background included into the term $C(E)$ of the equation (1)
contains at least 4 components:

1. Excitation of the matter of detectors and surrounding material
by energetic protons, and subsequent radiative decay of excited
isotopes (line-like emission). This decay leads to the retarded
increase of the background with a characteristic time, determined
by the half-decay times of  the excited isotopes.

2. The intrinsic background induced by   charged particles
(electrons and nuclei) instantly interacting with  detectors
(continuum emission).

3. The true CXB from the regions on the sky never occulted by the
Earth.

4. Resolved and unresolved Galactic and extragalactic sources.

\subsection{Radiation belts}

The charged particles at the INTEGRAL are monitored by the IREM
instrument (Hajdas et al., 2003). According to the IREM, the
satellite crosses the outer radiation belt containing electrons,
up to the distance $\sim 7 R_\oplus$, during 7 - 10 hours. Along
the orbit the flux of electron with energies $E_e
>0.5$ MeV is decreasing during this time from $F_e=10^7\,\,{\rm
cm}^{-2}{\rm s}^{-1}$ up to the unperturbed flux  $F_e=500\,\,{\rm
cm}^{-2}{\rm s}^{-1}$.  The inner radiation belt consists of
protons. As the satellite crosses this belt during $\sim 1.5$
hours,  the proton density is decreased $\sim 5$ times, from
$F_p=10\,\,{\rm cm}^{-2}{\rm s}^{-1}$   down to  $F_p=2\,\,{\rm
cm}^{-2}{\rm s}^{-1}$ in the unperturbed state (Hajdas et al.,
2003). Obviously the influence of the inner radiation belt on the
intrinsic background results mostly from the isotope decay with a
half-decay time of the order of  few hours.

As we will see from the consideration of SPI-ACS count rate (sect.
3.3), the influence of the internal radiation belt on the
background continuum is negligible. Indeed the SPI-ACS background
variations do not possess a regular decay pattern after the belt
passing. However the tail of the outer radiation belt may
influence on the detectors far away (see sect. 3.3)

\subsection{SPI and SPI/ACS background}

The background variations for the SPI and SPI/ACS instruments have
been investigated by Jean et al. (2003). It was shown   that the
outer electron belt is influencing  the continuum background
considerably, but it does not excite the isotopes. These
variations influence mainly the background at energies $<$100~keV.
The measurements presented by Jean et al. (2003) show $\sim 10\%$
variations arising from the 27 days solar rotation period.
Periodic variations correlated with the $\sim 3$ days period of
the satellite revolution around the Earth are also visible.  Their
amplitude is
 estimated as $\sim 2\%$. The authors (Jean et al., 2003)
conclude that "The total background variations are weak (less than
a few percent) thanks to the high eccentricity orbit and high
perigee of the INTEGRAL observatory."

Due to the Anti Coincidence Shield  (ACS) the absolute background
level in the Ge detectors of SPI is significantly depressed.
However the relative background variations are similar in both SPI
and SPI/ACS (see fig. 6 of Jean et al., 2003) and the SPI/ACS
background variations can be a tracer of the variations in the Ge
detectors of SPI in a wide energy band.

Particular cases of background variations in SPI-ACS during CXB
observations are presented in Figure 4. Non-monotonous SPI-ACS
background variations  can be explained by the different geometry
of the INTEGRAL orbit against the Earth's magnetopause and
magnetotail due to the orbit precession. To the author's knowledge
currently  there is no exact model of the SPI-ACS background
evolution.

\begin{figure}
\centerline{ \psfig{figure=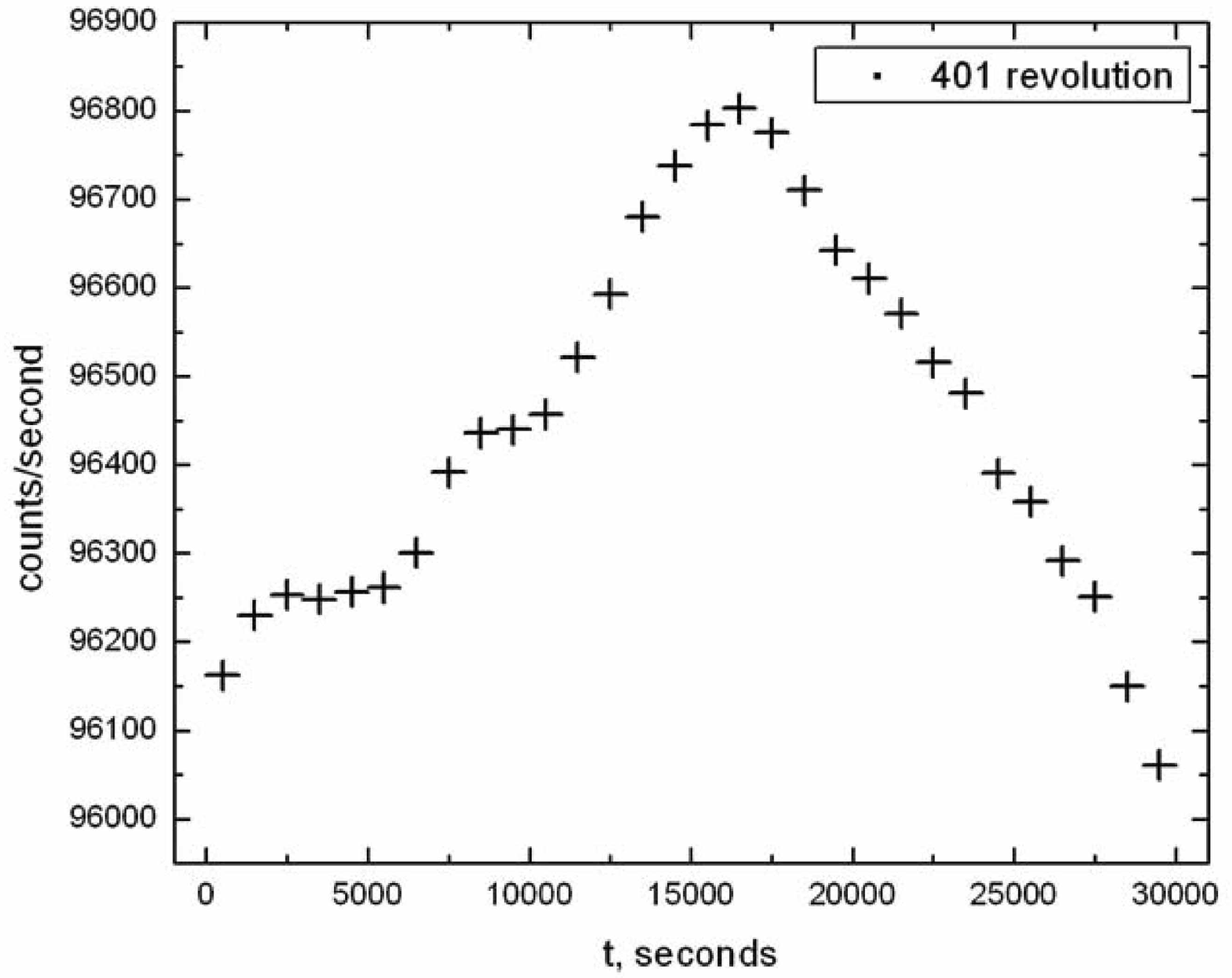,height=2.5in}
              \psfig{figure=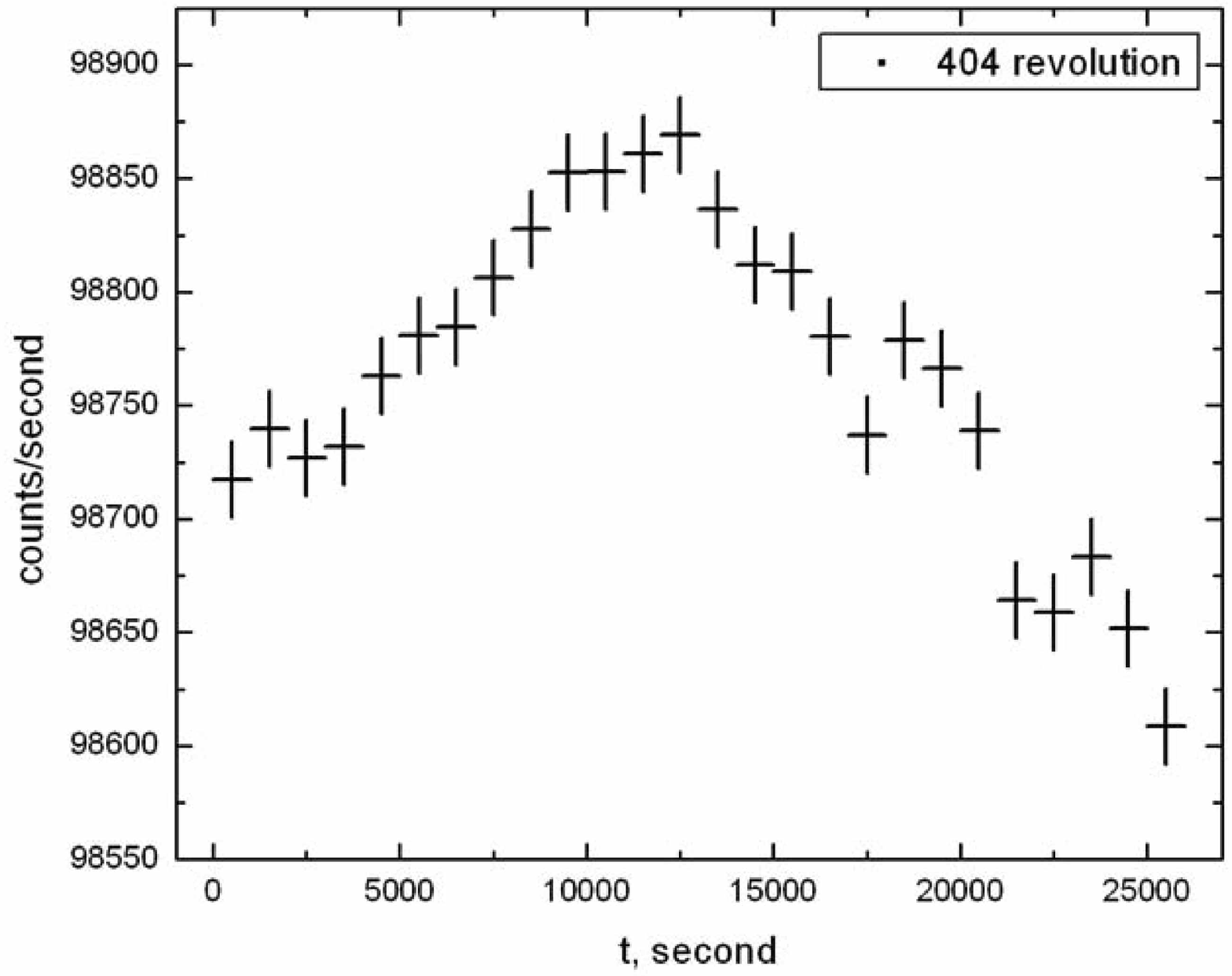,height=2.5in}}
\centerline{ \psfig {figure=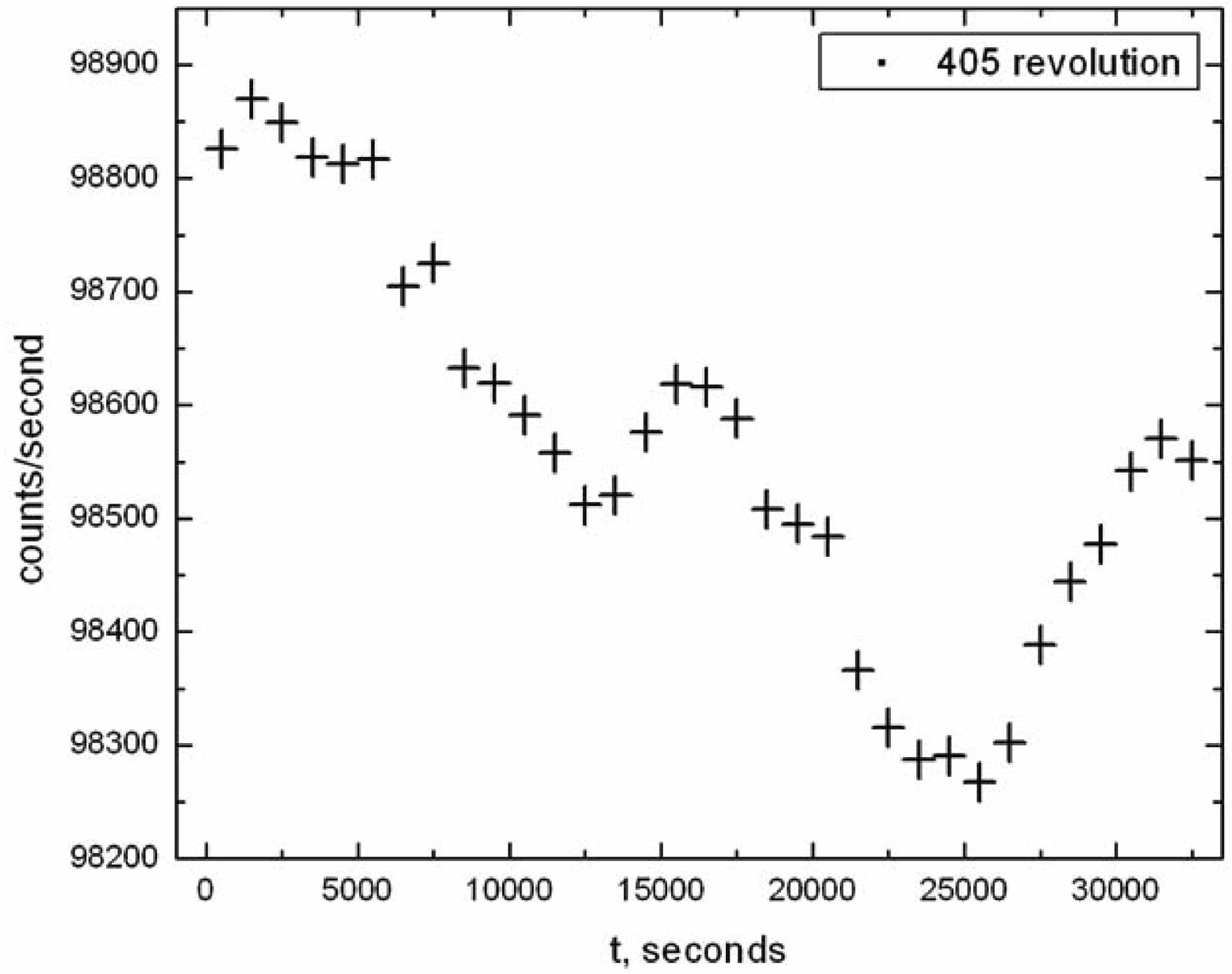,height=2.5in}
              \psfig {figure=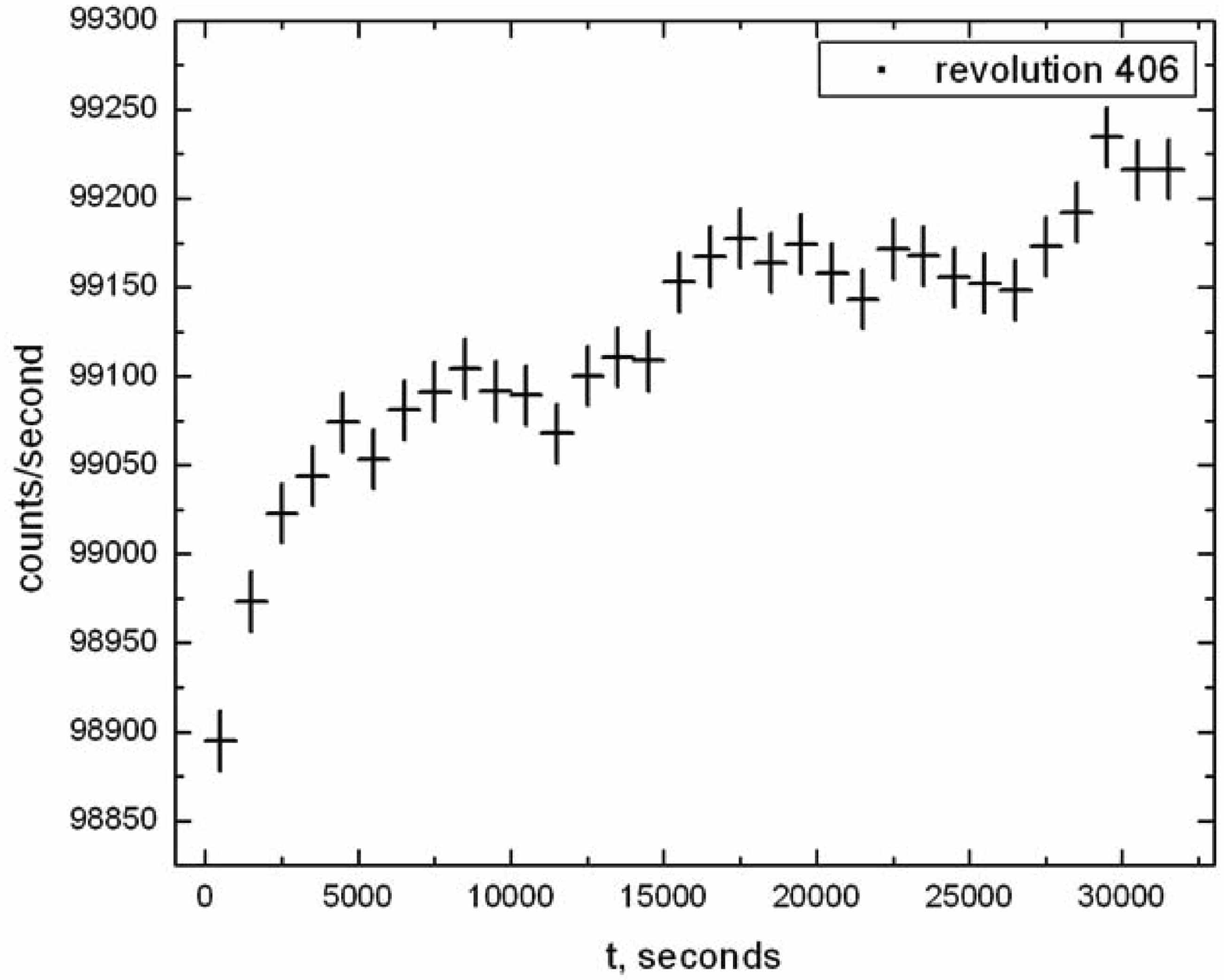,height=2.5in}}
\caption{Long term SPI-ACS background evolution during CXB
measurements in revolutions 401, 405 (left column), and 404, 406
(right column). Initial data of 50~ms resolution is rebinned in
1000~s. Null in the X-axis corresponds to the switch on time of
the SPI-ACS after the Earth radiation belt crossing. }
\label{spi-acs}
\end{figure}

\subsection{JEM-X background}

The signature of the variability of the internal background can be
suggested in Fig.3 of CE2007, where the level of the $C(E)$ term
before the Earth occultation is larger than after it, according to
the measurements of the JEM-X. Partially it might be attributed to
a stability of JEM-X gain after turning on  the instrument (see
section 2.3 of CE2007). Alternatively the variability can be
related to the intrinsic instrument background due to changing of
the radiation conditions.

The background in  JEM-X was discussed by Houvelin et al. (2003).
It was concluded that "the total background level varied with a
range of approximately 5\% between different orbital sections. the
variation is significant, but small."

\subsection{IBIS/ISGRI background}

The IBIS/ISGRI sensitivity and background is discussed in the
paper of Lebrun et al. (2005).  However, to the author' knowledge,
there are no investigations of the intrinsic ISGRI detector
background along the INTEGRAL orbit. It was suggested by the
authors of TM2010 that the intrinsic ISGRI background correlates
with the SPI-ACS count rate. While in general this is a quite
reasonable suggestion confirmed by the investigation of
correlation of the count rate in IBIS/ISGRI and SPI-ACS during
solar flare (see details in TM2010), one can take into account
that spallation reactions of different chemical elements in
detectors of SPI-ACS (Ge, Bi) and ISGRI (Cd, Te) may lead to a
difference in the background evolution in particular energy bands.

\section{Influence of the background variations on the accuracy of CXB measurement }

One can see that  the amplitudes of the intrinsic background
variations along the orbit are specific for each instrument and
the energy band, and can vary from 2$\%$ (sect. 3.3) up to 5$\%$
(sect. 3.4). These variations should be taken into account in the
long term observations, if the required accuracy  is comparable
with the amplitude of the background variation. In the particular
case of CXB measurement the amplitude of the intrinsic background
variations should be compared  with amplitude of the CXB
modulation related to the Earth occultation.

The CXB modulation is maximal in energy band 20-40 keV, where the
energy spectrum of CXB has the maximum (see Fig. 2). In this band
the total effect (CXB depression due to the Earth occultation plus
the Earth atmosphere emission and the Earth CXB reflection (Earth
albedo) can be estimated as much as $\sim$ 10 $\%$ of the total
flux of CXB in FOV registered in SPI and IBIS/ISGRI (see e.g Fig.
7 of CE2007). The uncertainty induced by the background variation
during the CXB measurement can be estimated as a ratio of the
amplitude of the background variation to the amplitude of the CXB
modulation. The actual amplitude of the background variation
during the Earth occultation observation can be estimated as low
as 1\% (see Fig. 4). As we have seen above (sect. 3.3) the
background in SPI-ACS is a tracer of the background in SPI
detectors. Thus, the uncertainty of the CXB measurement induced by
the background variation in SPI is equal to 1/10 = 10\%, provided
the flux corresponding to the intrinsic background is comparable
to the flux of CXB in the aperture of the telescope. At least for
SPI and IBIS/ISGRI the intrinsic background is comparable with CXB
component in the total count rate.

The uncertainty induced by the background variations is increasing
beyond the peak of the CXB spectrum because the CXB modulation is
decreasing (see e.g. Ajello et al., 2008).

Accounting of the intrinsic background is an essential part of the
accuracy improvement of the CXB measurements.  Modelling of the
intrinsic background in the CXB occultation technique is  used in
the latest CXB observations (e.g. Ajello et al., 2008) and new
analysis of IBIS/ISGRI data (TM2010) of the Earth occultation
observations in 2006 (CE2007).

\section{CXB variations due to unresolved sources}

The CXB flux measured in the telescope aperture varies due to
 a number of unresolved compact sources, both  Galactic and
extragalactic. Indeed, we do not know \textit{a priori} which
unresolved source belongs to the Galaxy, and which one is
extragalactic. Poisson variations of the source number in FOV at
the sensitivity of observation, lead to the uncertainty in the CXB
measurement. While the sensitivity of the IBIS/ISGRI in 20-50 keV
band is $f_x \sim 10^{-11}\, {\rm ergs}\, \, {\rm cm}^{-2}\, {\rm
s}^{-1} $ (i.e. the sensitivity in a deep extragalactic survey
(Krivonos et. al, 2005) the actual sensitivity during the CXB
measurement can be estimated as $\sim 4.5\cdot10^{-11} \, {\rm
ergs}\, \, {\rm cm}^{-2}\, {\rm s}^{-1} $. This sensitivity
roughly corresponds the faintest source detected in FOV of
IBIS/ISGRI (see the Table 2 of CE2007). Using
$[\log(N(>f))=-(3/2)\log(f)]$ cumulative distribution for the
extragalactic sources (Krivonos et. al, 2005) one can obtain
normalization of the distribution at the actual sensitivity flux
level $\sim 4.5\cdot10^{-11} \, {\rm ergs}\, \, {\rm cm}^{-2}\,
{\rm s}^{-1} $ as $1.4\cdot10^{-3}\, {\rm deg^{-2}}$, which leads
to the uncertainty of CXB measurements of about $\sim$2\%.

Estimates of the uncertainty due to the unresolved galactic
sources is less strict due to the presence the Galactic Ridge as
well as numerous galactic sources in FOV of telescopes. The
normalization of cumulative distribution
$[\log(N(>f))=-(1)\log(f)]$ can be estimated as the number of the
sources detected during the CXB observation (equal to 11) divided
by the effective FOV of IBIS/ISGRI ($\sim$ 15x15 deg), resulting
to $0.05 \,\,{\rm deg^{-2}}$. Thus, the the number of the galactic
sources in FOV is 35 times  larger than the number of the
extragalactic sources, and the uncertainty of CXB measurement due
to the unresolved galactic sources is negligible.

\section{Normalization of the Earth atmosphere emission}

The flux of the Earth atmosphere emission induced by cosmic rays
is comparable to the CXB component, while the Earth disk in FOV of
the INTEGRAL telescopes. The Earth atmosphere emission was
calculated  in the paper of Sazonov et al. (2007). It was shown
that the photon spectrum of the Earth atmosphere emission peaks at
about 44~keV, and the normalization of the approximation law is
equal to $31.7 \, {\rm keV^{2}} {\rm cm^{-2}} {\rm s^{-1}} {\rm
keV^{-1}} {\rm sr^{-1}}$ (c.f. the observed flux near the maximum
of the  spectrum of CXB at 29~keV is equal to $47 \, {\rm keV^{2}}
{\rm cm^{-2}} {\rm s^{-1}} {\rm keV^{-1}} {\rm sr^{-1}}$). Indeed
the uncertainty of the Earth atmosphere emission should be
included in the total uncertainty of the CXB measurement by the
Earth occultation technique. In the procedure of CXB calculation
(see equation (1), the normalization factors of both the Earth
emission component, and CXB were free parameters, and the
normalization of the Earth atmosphere emission was estimated as
$32.9\pm1.3 \, {\rm keV^{2}} {\rm cm^{-2}} {\rm s^{-1}} {\rm
keV^{-1}} {\rm sr^{-1}} $. It is well within 1 sigma confidence
level of the predicted value (31.7) and leads to the $\sim$4\%
uncertainty of the Earth emission component determination.
Therefore uncertainty of about 4\% is introduced   by the Earth
atmosphere emission in the CXB measurements by   the Earth
occultation technique.

\section{Conclusion}

Summing up the uncertainties of the CXB measurement by the Earth
occultation
 technique around CXB peak (20 - 40~keV) in INTEGRAL observations (CE2007), one can
 consider at least the following sources of
uncertainties. The compact source reflection came as $\sim3\%$.
The intrinsic background variability along the orbit can
contribute as much as 10\%. Variations due to the unresolved
sources at the limiting sensitivity is $\sim$2\%. Modelling of the
atmosphere reflection $\sim 4\%$. We should add these values to
the values already discussed and estimated in the paper CE2007.
Those include
 statistical error in the normalization of the CXB component (joint
fit to JEM-X, IBIS/ISGRI, and SPI data) as $\sim 1\%$; neglecting
contribution of the compact source 4U1626-67 in FOV as $\leq 2\%$;
uncertainty of the Crab photon index as $\sim 1\%$ (the accuracy
of absolute calibrations of the instruments onboard the INTEGRAL
observatory is also  estimated as low as $\sim 1\%$ in Jourdain et
al., (2008). Combining all these uncertainties we obtain the total
uncertainty in the 20-40~keV energy band as $\sim 11\%$. One can
see that the main uncertainty of the CXB measurement of the
INTEGRAL observatory is induced by the intrinsic background
variations.

A comprehensive consideration of the CXB accuracy induced by the
absolute calibration, based on Crab observation in different
experiments, can be found in CE2007.

Considering  the precision of the recent measurements at the CXB
peak $\sim 30$~keV, (CE2007, Frontera et al., 2007, Ajello et al.,
2008)   one can see that practically all measurements are
consistent within their uncertainties (see also Moretti et al.,
(2009)) and are comparable with the HEAO-1 observations (Gruber et
al., 1999).

Note that estimates of CXB around its peak in recent BeppoSAX
observation (Frontera et al. 2007; see also Frontera et al. 2008)
and re-analysis of CXB obtained with the IBIS/ISGRI of the
INTEGRAL (TM2010) show a good agreement with the historic HEAO-1
measurements. Meanwhile our consideration shows that  the CXB
normalization obtained at least for the SPI and IBIS/ISGRI
detectors in CE2007 can be also compatible with the HEAO-1
measurements if we account the intrinsic background variation
during the CXB measurement. We stress that modelling of the
intrinsic background is necessary for increasing accuracy  of any
long term observations performed by the INTEGRAL observatory, and
in particular in the CXB measurements.

We have not discussed several other sources of uncertainties
arising in the occultation technique. These are the Earth
atmosphere self-activity, such as Auroral emission, Terrestrial
Gamma-ray Flashes; intrinsic variability of the X-ray sources in
FOV and transient phenomena, such as Gamma-Ray Bursts, SGR, and
X-ray bursters etc. While influence of some of them on the CXB
measurements may be non negligible the investigation of these
source is beyond the scope of this paper.

Despite decreasing statistical uncertainties in the near Earth CXB
observations systematic uncertainties continue to dominate. One
possible way toward improving the accuracy of  the measurement
around the CXB peak is to use the lunar orbital spacecraft and
Moon as a natural block and CXB modulator. Indeed the Moon is a
more perfect block with absence of any self activity (such as
auroral Earth emission), and having the lowest albedo in CXB peak
spectrum (Churazov et al., 2008). Moreover the absence of the
magnetosphere make the lunar orbital experiments (e.g. past Lunar
Prospector Gamma Ray Spectrometer, Feldman et al., 1999) suitable
for the precise CXB measurement. Meanwhile, the observation of the
dark side of the Moon by Chandra observatory is already used in
the CXB measurement in 2-7~keV energy range (Markevitch et al.,
2002).

\section*{REFERENCES}

\noindent
Ajello M., Greiner J., Sato G. , et
al. 2008, ApJ    689, 666

\noindent
Churazov E., Sunyaev R., Revnivtsev M., et al. 2007,  A\&A, 467,
529

\noindent  
Churazov E., Sazonov S., Sunyaev R., Revnivtsev M.
2008, MNRAS    385 , 719

\noindent 
Feldman W.C., Barraclough B.L., Fuller K.R., et al.
1999,
Nuclear Instruments and Methods in Physics Research A,   422, 562

\noindent 
Frontera F., Orlandini M., Landi R., et al. 2007, ApJ,  666, 86

\noindent 
Frontera F., Orlandini M., Landi R., et al. 2008,
Chinese Journal of Astronomy \& Astrophysics Supplement,  8, 297

\noindent 
Fukada, Y., Hayakawa, S., Kasahara, I.,  et al.
1975, Nature, 254, 398

\noindent 
Giacconi R., Gursky H., Paolini F. R., Rossi B. B.
1962, Physical Review Letters   9 , 439

\noindent
Gilli R., Comsatri A., Hasinger G., 2007, A\&A, 463, 79

\noindent  
 Gruber D. E., Matteson J. L., Peterson L. E., Jung,
 G.V. 1999, ApJ,  520, 124

\noindent 
Hajdas W., B\"uhler P.,  Eggel C., F et al. 2003,  A\&A 411, L43

\noindent 
Harmon B.A., Wilson C.A., and Fishman G.J. 2004, ApJ Suppl.  154,
585

\noindent 
Huovelin J., Maisala S., Schultz J.,  et al. 2003,  A\&A   411 , L253

\noindent 
Jean P., Vedrenne G., Roques J.P., et al.  2003,  A\&A  411,
L107

\noindent 
Krivonos R., Vikhlini A., Churazov E., et al,
2005, ApJ,  625, 89

\noindent 
Lebrun F., Roquesc J.-P., Sauvageon A.,  et al., 2005,
Nuclear Instruments and Methods in Physics Research A 541, 323

\noindent   
Markevitch M., Bautz M. W., Biller B., et
al, 2003, ApJ,  583, 70

\noindent   
Mazets E.P., Golenetskii S.V., Il'inskii V.N.,
et al.  1974, JETP Letters,  20 , 32.

\noindent 
Moretti  A., Pagani  C., Cusumano G., et
al. 2009, A\&A  493, 501

\noindent   
Sazonov S., Churazov E., Sunyaev R., Revnivtsev M.
2007, MNRAS   377, 1726

\noindent
Sazonov S., Krivonos R., Revnivtsev M., et al. 2008, A\&A, 482,
517


\noindent
Treister E., Urry C.M., Virani S., 2009, ApJ, 696, 110

\noindent
T\"{u}rler M., Chernyakova M., Courvoisier T.J.-L., et al. 2010,
arXiv:1001.2110 [astro-ph.CO]

\noindent
Ueda Y., Akiyama M., Ohta K., et al. 2003, ApJ, 598, 886

\noindent 
Verrecchia F., in 't Zand J. J.M., Giommi1 P., et al.
2007,  A\&A  472, 705

\noindent 
Winkler C.,  Courvoisier T. J.-L., Di Cocco G., et
al. 2003, A\&A 411, L1

\end{document}